\documentclass[aps,showpacs,PRL,twocolumn,superscriptaddress]{revtex4-2}
\makeatletter

\newcommand{\Rmnum}[1]{\expandafter\@slowromancap\romannumeral #1@}
\makeatother
\usepackage[english]{babel}
\usepackage{amsmath}
\usepackage{graphicx}
\usepackage{subfigure}
\usepackage{color}\usepackage[colorlinks=true,citecolor=blue,urlcolor=blue,linkcolor=blue,hyperindex]{hyperref}
\usepackage{amsfonts}
\usepackage{tikz}
\usepackage{esint}
\usepackage{mathrsfs}
\usepackage{verbatim}
\usepackage[normalem]{ulem}
\usepackage{titlesec}
\usepackage{appendix}

\newcommand{\bea}{\begin{eqnarray}}
\newcommand{\eea}{\end{eqnarray}}

\titleformat{\subsection}[runin]
  {\normalfont\itshape} 
  {\thesubsection}{1.0em} {} 
\titleformat{\section}[runin]
{\normalfont\bfseries\itshape}
  {\thesection}{0.1em} {} 
\begin{document}
\title{Spectroscopic evidences for the spontaneous symmetry breaking at the $SO(5)$ deconfined critical point of $J$-$Q_3$ model}
\author{Shutao Liu}
\affiliation{Department of Physics and State Key Laboratory of Surface Physics, Fudan University, Shanghai 200438, China}

\author{Yan Liu}
\affiliation{Department of Physics and State Key Laboratory of Surface Physics, Fudan University, Shanghai 200438, China}

\author{Chengkang Zhou}
\affiliation{Department of Physics and HKU-UCAS Joint Institute of Theoretical and Computational Physics, The University of Hong Kong, Pokfulam Road, Hong Kong SAR, China}

\author{Zhe Wang}
\affiliation{Department of Physics, School of Science and Research Center for Industries of the Future, Westlake University, Hangzhou 310030,  China}
\affiliation{Institute of Natural Sciences, Westlake Institute for Advanced Study, Hangzhou 310024, China}

\author{Jie Lou}
\affiliation{Department of Physics and State Key Laboratory of Surface Physics, Fudan University, Shanghai 200438, China}
\affiliation{Collaborative Innovation Center of Advanced Microstructures, Nanjing 210093, China}

\author{Changle Liu}
\email{liuchangle89@gmail.com}
\affiliation{School of Physics and Mechatronic Engineering, Guizhou Minzu University, Guiyang 550025, China}

\author{Zheng Yan}
\email{zhengyan@westlake.edu.cn}
\affiliation{Department of Physics, School of Science and Research Center for Industries of the Future, Westlake University, Hangzhou 310030,  China}
\affiliation{Institute of Natural Sciences, Westlake Institute for Advanced Study, Hangzhou 310024, China}

\author{Yan Chen}
\email{yanchen99@fudan.edu.cn}
\affiliation{Department of Physics and State Key Laboratory of Surface Physics, Fudan University, Shanghai 200438, China}
\affiliation{Collaborative Innovation Center of Advanced Microstructures, Nanjing 210093, China}

\date{\today}
\begin{abstract}
{Recent numerical and theoretical studies on the two-dimensional $J$-$Q_3$ model suggests that the deconfined quantum critical point is actually a $SO(5)$-symmetry-enhanced first-order phase transition that is spontaneously broken to $O(4)$. However, this conclusion has mainly relied on finite-size scaling of the entanglement entropy, lacking direct evidence from physical observables.} Here, we investigate the dynamical spectra of spin and bond operators at the deconfined critical point of the $J$-$Q_3$ model using large-scale quantum Monte Carlo simulations, and contrasting them with the well-established $\mathrm{O(3)}$ Wilson-Fisher criticality in the $J_1$-$J_2$ Heisenberg model. Although both models exhibit two gapless magnon modes in the N\'eel phase, their critical behaviors diverge strikingly. At the $J_1$-$J_2$ critical point, the Higgs mode becomes gapless, yielding three gapless modes that reflect the full restoration of the $\mathrm{O(3)}$ symmetry. 
{In the $J$-$Q_3$ model, we instead observe four gapless transverse modes at the either side of the transition. 
This spectral feature, together with the entanglement entropy results, provides direct evidence for the weakly first-order scenario that the deconfined quantum critical point exhibits an emergent $\mathrm{SO(5)}$ symmetry that spontaneously breaks to $\mathrm{O(4)}$.}
\end{abstract}

\maketitle

\textit{\color{blue}Introduction.---}
The deconfined quantum critical point (DQCP) has attracted significant attention since decades ago~\cite{senthil2004deconfined,senthil2004quantum,sandvik2007evidence,nahum2015deconfined,
qin2017duality,wang2017deconfined}, and is believed to offer a new paradigm going beyond the Laudau-Ginzburg-Wilson framework of phase transitions. 
The DQCP scenario predicts that the phase transitions from the collinear N\'eel antiferromagnet (AFM) to the valence-bond solids (VBS) in 2D magnetic systems can be continuous, despite the incompatible broken symmetries.
A. Sandvik proposed the sign-free spin-1/2 $J$-$Q$ model on a square lattice that realizes this N\'eel-VBS transition, allowing us to study the DQCP phenomena by large-scale quantum Monte Carlo (QMC) simulations~\cite{sandvik2007evidence,lou2009antiferromagnetic}.Numerical simulations in the ground state~\cite{sandvik2007evidence,lou2009antiferromagnetic,sandvik2010continuous}
and at finite temperatures~\cite{melko2008scaling} show continuous transitions that seems to be
consistent with the prediction of the DQCP.

The DQCP scenario of the N\'eel-to-VBS transition conjectures that this transition is continuous described by a conformal field theory (CFT) with emergent SO(5) symmetry out of the O(3)$\times\mathbb{Z}_4$ microscopic Hamiltonian. In this picture, the three-component O(3) N\'eel order parameter and the two-component $\mathbb{Z}_4$ VBS order parameter combine into a five-component vector under an enlarged $\rm{SO(5)}$ symmetry. 
The presence of this emergent $\rm{SO(5)}$ symmetry has been confirmed by a series of later studies~\cite{nahum2015emergent,wang2017deconfined,sreejith2019emergent,zhao2019symmetry,takahashi2020valence}.
However, the nature of such transition still remains highly debated. 
Several large-scale numerical simulations indicate a drift of the critical exponents with increasing system sizes, casting doubt on whether the transition is truly continuous or weakly first-order in the thermodynamic limit. 
Moreover, applications of the non-perturbative conformal bootstrap with this $\rm{SO(5)}$ CFT reveal that critical exponents numerically measured in the previous studies are beyond the strict bounds imposed by conformal symmetry~\cite{nakayama2016necessary,li2022bootstrapping,poland2019conformal}. 
In addition, the recent QMC work on entanglement entropy (EE) scaling analysis reveals an anormal subleading-correction which violates the prediction of unitary CFT~\cite{zhao2022scaling,song2024extracting,song2025evolution,wang2025extracting,deng2024diagnosing}.

To reconcile the discrepancies between theoretical expectations and numerical findings, two main scenarios have been put forward, namely pseudo-criticality~\cite{zhou2024so,wang2017deconfined,nahum2015deconfined,nahum2020note,ma2020theory}
and multi-criticality~\cite{zhao2020multicritical,chen2024phases,lu2021self,chester2024bootstrapping}. 
In these scenarios, without further fine tuning such DQCP flows to a weakly first-order transition with a tiny gap and a large but finite correlation length. 
The scenario of weakly first-order transition with enhanced SO(5) symmetry is evidenced by the recent numerical EE study of the $J$-$Q_3$ model~\cite{deng2024diagnosing}. 
Through a finite size scaling analysis of the EE~\cite{deng2023improved}, they observe a negative logarithmic correction, whose coefficient corresponds to the presence of four Goldstone modes, 
indicating a spontaneous symmetry breaking from $\rm{SO(5)}$ into $\rm{O(4)}$ at the DQCP~\cite{deng2024diagnosing}. While this results successfully solve the doubt of the strange EE scaling at DQCP, the extra finite size correction for the coefficient of the logarithmic term of EE still needs more demonstration.

While the EE provides compelling evidence for a weakly first-order transition at the ``DQCP'', it remains crucial to seek further confirmation from independent probes
{, such as direct identification of the four Goldstone modes through spectroscopic measurements.}  
{In previous studies, the spectroscopic investigations have primarily focused on the spin excitation channel, which was used to capture the fractionalized spinon excitations and emergent symmetry in the easy-plane $J$-$Q$ model with emergent $\rm{O(4)}$ symmetry~\cite{ma2018dynamical,ma2019role}. However, to fully capture the four Goldstone modes at the ``DQCP'' with emergent $SO(5)$ symmetry, it is not sufficient to measure only the spin channel. Probing the bond spectral functions is also essential. This leaves a gap in the current understanding of the DQCP.} In this work, we address this gap and identify the existence of four Goldstone modes through spectroscopic analysis.

\textit{\color{blue}Model and method.---}
\begin{figure}[htp]
  \centering
  \includegraphics[width={.45\textwidth}]{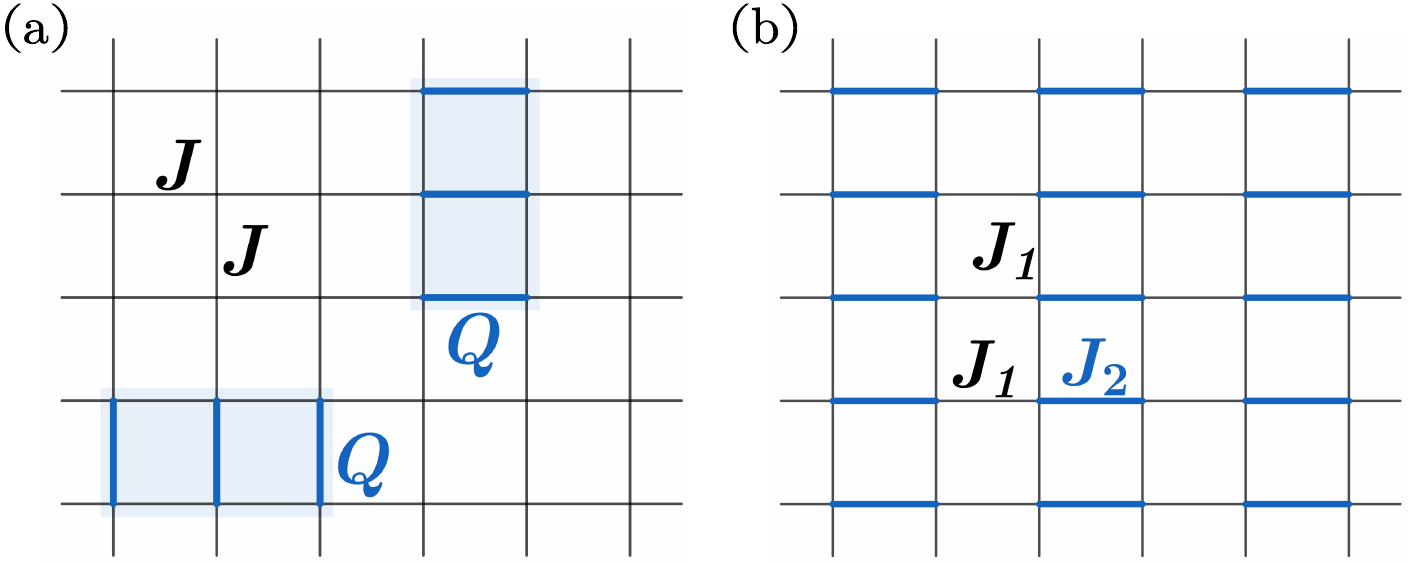}
  \caption{Illustration of the two lattice models: (a) the $J$-$Q_3$ model and (b) the $J_1$–$J_2$ antiferromagnetic model.}
  \label{fig:Hamiltonian}
\end{figure}
In order to study the low-energy behavior of the (2+1)$d$ system realizing a N\'eel-VBS transition, we investigate the $J$-$Q_3$ model described by the following Hamiltonian 
    \begin{equation}
      H_{J-Q_3}=-J\sum_{\left<ij \right>}P_{ij}-Q\sum_{\left<ijklmn\right>}P_{ij}P_{kl}P_{mn},
    \end{equation}
and $\left<ij \right>$ denotes the 
nearest-neighbor sites, and $\left<ijklmn \right>$ refers to 
the points of the column consisting six sites, such that $(i,j)$, $(k,l)$, $(m,n)$ form three 
adjacent parallel links (either vertical or horizontal) as showed in Figure \ref{fig:Hamiltonian} (a). $P_{ij}=\frac{1}{4} - \boldsymbol{S}_{i}\cdot \boldsymbol{S}_{j}$ is the singlet projector operator on sites $i$ and $j$,
where $\boldsymbol{S}_{i}$ denotes the spin-$1/2$ operator on the site $i$. 
For simplicity of our discussions, we 
define dimensionless coupling parameter $q = Q / (J+Q).$ 
For this $J$-$Q_3$ model, 
the system lies in the N\'{e}el phase for small $q$ and the $\mathbb{Z}_4$ VBS order for large $q$~\cite{lou2009antiferromagnetic}. 
The N\'eel-VBS phase transition occur at $q_c=0.59864(5)$~\cite{wang2022scaling}. 
The low-energy excitations are gapless magnons in the N\'eel phase~\cite{dalla2015fractional,song2023dynamical} and become gapped in the VBS phase. 
The order parameters of the N\'eel and the VBS phases are described by $\boldsymbol{N}=(N_x,N_y,N_z)$ and $(D_x,D_y)$ respectively. 
At DQCP, the emergent $\rm{SO(5)}$ symmetry of the N\'eel and the VBS order parameters has been observed in numerical studies~\cite{nahum2015emergent,takahashi2020valence,takahashi2024so}.
Although some studies of the $J$–$Q$ model have reported scaling violations, which have been interpreted as evidence for a weak first-order transition~\cite{jiang2008antiferromagnet,kuklov2008deconfined,chen2013deconfined,wang2025probing}, we follow the convention of referring to the transition point as DQCP.

For convenience of discussions, we combine the N\'eel and VBS orders into a five-component order parameter $\boldsymbol\phi=(N_x,N_y,N_z,D_x,D_y)^T$ that carries vector representation of the SO(5) symmetry. The generators of this SO(5) symmetry are denoted by $L^{\alpha\beta}=-L^{\beta\alpha}$ ($\alpha$, $\beta=1,2,3,4,5$ are the labels of the component)  which rotates between $\phi^\alpha$ and $\phi^\beta$.  
{In a symmetry-broken state, the low-energy excitations constitutes one Higgs mode as the longitudinal (amplitude) fluctuation of the order parameter $\phi^\alpha$ itself, as well as transverse fluctuations  $L^{\alpha\beta}$  that rotates the ordered component $\alpha$ to the disordered ones $\beta\neq\alpha$. If the   $L^{\alpha\beta}$ becomes the symmetry generator of the system, such transverse mode will become gapless and are also known as ``Goldstone'' modes. }


To make a comparison with the $J$-$Q_3$ model, we also study a square lattice columnar dimerized $J_1$-$J_2$ AFM Heisenberg model described by the Hamiltonian~\cite{matsumoto2001ground,wenzel2009comprehensive}
\begin{equation}
  H_{J_1-J_2}=J_1\sum_{\left<ij \right>_1}\boldsymbol{S}_i\cdot \boldsymbol{S}_j+J_2\sum_{\left<ij \right>_2 }\boldsymbol{S}_i\cdot \boldsymbol{S}_j,
\end{equation}
where $\left<ij \right>_1$ and $\left<ij \right>_2 $ denotes the thin bond and the thick bond respectively as shown in Figure \ref{fig:Hamiltonian}(b). We set the exchange energy of thick bonds greater than the thin ones, i.e., $g=J_2/J_1>1$. 
The phase diagram of this model contains two phases. 
For large $g$, the system lies in the trivial dimerized phase that preserves the O(3) spin-rotational symmetry. The low-energy spin excitations are spin triplets corresponding to fluctuations of the three $\boldsymbol{N}$ components. 
For small $g$, the system lies in the N\'eel phase with $\langle\boldsymbol{N}\rangle\neq0$ that spontaneously breaks the O(3) symmetry. The low-energy spin excitations are a pair of Goldstone modes that dictate the O(3) spontaneous symmetry breaking to O(2), in addition to a gapped Higgs mode that characterizes the amplitude fluctuations of $\boldsymbol{N}$.
The N\'eel and the dimerized phases are separated by a $(2+1)d\ \rm{O(3)}$ quantum phase transition (QPT) that locates at $g_c=(J_2/J_1)_{c}=1.90951(1)$~\cite{matsumoto2001ground,wenzel2009comprehensive,ma2018anomalous,chakravarty1988low,haldane19883,chubukov1994theory,wenzel2008evidence}. 
\begin{figure}[!t]
  \centering
  \includegraphics[width=0.45\textwidth]{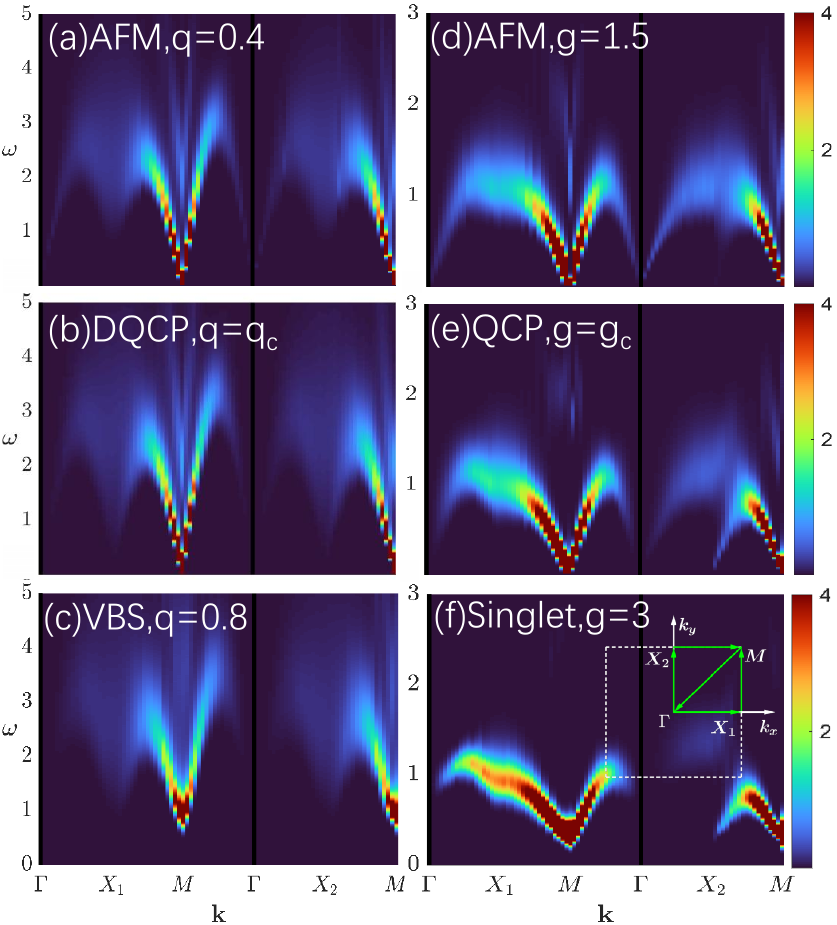}
  \caption{Spin spectral functions $A_s(\boldsymbol{k},\omega)$ obtained from QMC-SAC.
  (a),(b),(c) show the spectral function of $J$-$Q_{3}$ model 
  in N\'{e}el phase ($q=0.4$), at the DQCP ($q=q_c$) and in the VBS phase ($q=0.8$) respectively.
  (d),(e),(f) show the spectral function of $J_1$-$J_2$ model in 
  the N\'{e}el phase ($g=1.5$), near the QCP ($g=g_c$) and in the dimer state ($g=3.0$) respectively.~\cite{commutator}}
  \label{spin}
\end{figure}

To investigate the quantum dynamics of the systems, we perform large-scale QMC simulation using the stochastic series expansion (SSE) technique to simulate these two models~\cite{sandvik1991quantum,Sandvik1999SSE,syljuaasen2002quantum,sandvik2019stochastic,yan2019Dimer,yan2021DimerImproved,desai2021resummation}. 
For completeness, both spin-spin and dimer-dimer correlations are measured in our study.
The imaginary time spin-spin correlation is expressed as  $  G_{s}(\boldsymbol{k},\tau)= \frac{1}{L^2}\sum_{i,j}e^{-i\boldsymbol{k}\cdot (\boldsymbol{r}_i-\boldsymbol{r}_j) } \left<\boldsymbol{S}_{i}(\tau)\cdot\boldsymbol{S}_{j}(0) \right> $.
Similarly, we also measure the dimer correlation of the $x$-bonds ~\cite{zhou2021amplitude,dorneich2001accessing}, as given by $ G_b(\boldsymbol{k},\tau)=\frac{1}{L^{2}}\sum_{i,j}e^{-i \boldsymbol{k}\cdot \left( \boldsymbol{r}_i - \boldsymbol{r}_j  \right)}\left<B_{x,i}(\tau)B_{x,j}(0) \right>$,
where $B_{x,i}=\boldsymbol{S}_i\cdot\boldsymbol{S}_{i+\hat{x}}$ is a spin singlet bond operator along the $x$-direction.
The imaginary time correlation functions $G_{s,b}(\tau)$ are related to the real frequency spectral function $A_{s,b}(\omega)$ by
$G_{s,b}(\tau)=\frac{1}{\pi}\int_{0}^{\infty}\mathrm{d}\omega A_{s,b}(\omega)e^{-\tau \omega}.$ 
The frequency-domain spectral functions are extracted from imaginary time correlations by performing stochastic analytic continuation (SAC) 
\cite{sandvik1998stochastic,sandvik2016constrained,beach2004identifying,yan2021topological,zhou2021amplitude,shao2023progress}.
\begin{figure}[!htbp]
  \centering
  \includegraphics[width=0.45\textwidth]{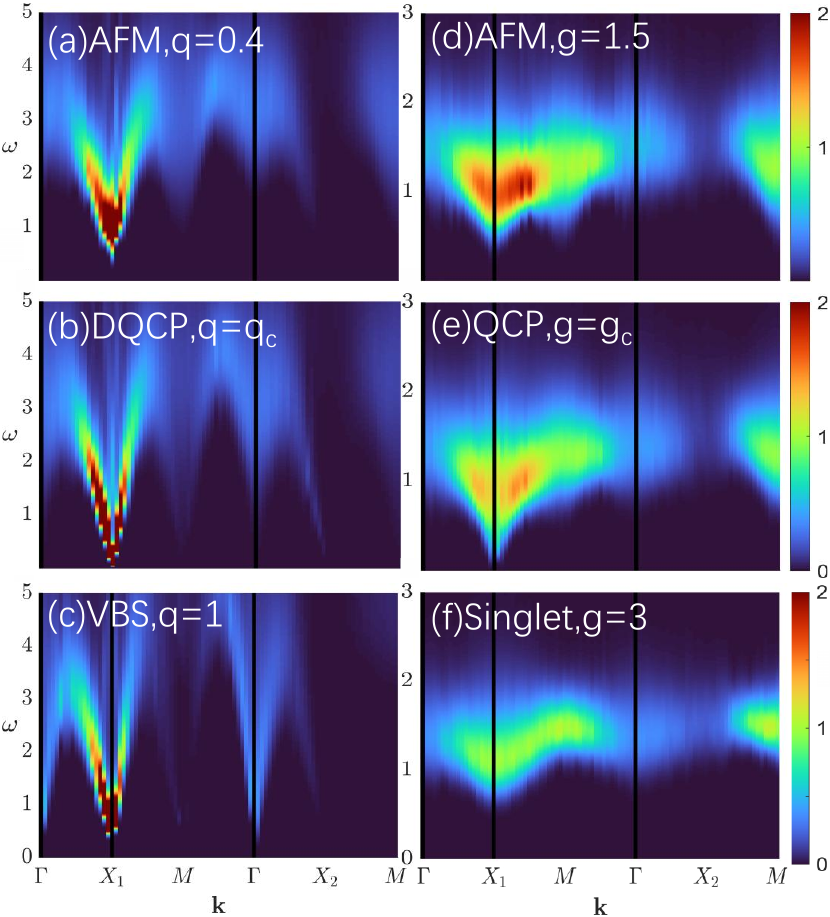}
  \caption{Bond spectral functions $A_b(\boldsymbol{k},\omega)$ obtained from QMC-SAC. (a),(b),(c) show the spectral function of $J$-$Q_3$ model in the N\'{e}el phase ($q=0.4$) near the DQCP ($q=q_c$) and in the VBS phase ($q=1.0$) respectively. (d),(e),(f) show the spectral function of $J_1$-$J_2$ model in the N\'{e}el phase ($g=1.5$), near the QCP ($g=g_c$) and in the dimer state ($g=3.0$) respectively~\cite{bond_commutator}. 
  }
  \label{bond}
\end{figure}

\textit{\color{blue}N\'eel AFM phase.---}
We present numerical results of spectral functions of spin correlations in Fig. \ref{spin} and bond correlations in Fig. \ref{bond} of 
the $J$-$Q_3$ model and the $J_1$-$J_2$ model.
The calculated system has periodic boundary condition with length $L=36$.
The QMC calculations are carried out at inverse temperature $\beta=4L$ along the high-symmetry path in the Brillouin zone (BZ) showed in the Fig. \ref{spin} (f). 

\begin{figure}[thbp]
\centering
\includegraphics[width=0.45\textwidth]{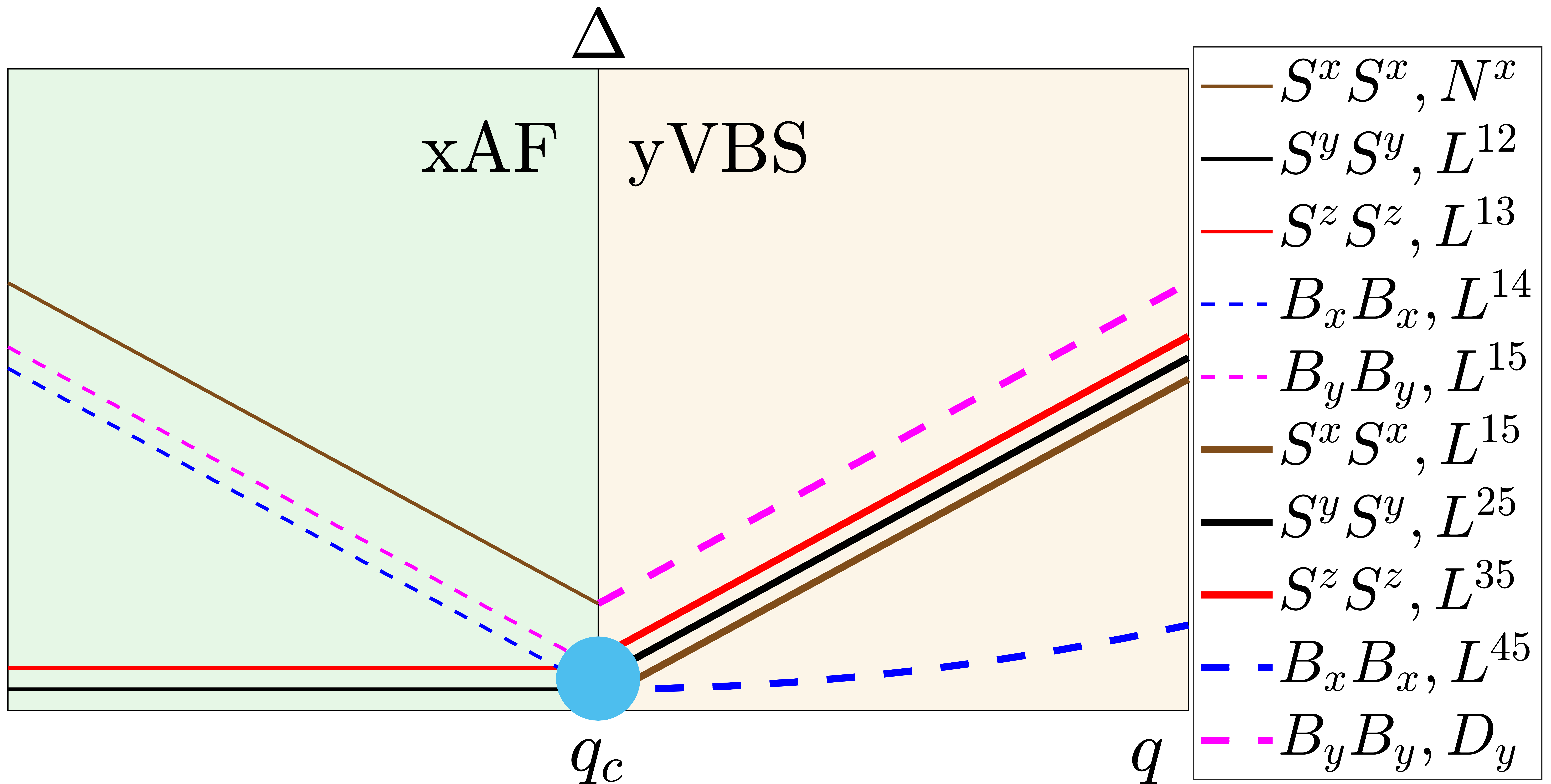}
\caption{Schematic spectral evolution for the SO(5) symmetry-enriched first order scenario of the N\'eel-VBS transition \cite{chen2024spin}. xAF means the system is in the N\'eel phase and breaks into x direction in spin space. yVBS means the system is in the VBS phase and breaks into the y direction in VBS order parameter space. 
The nature of modes and the correlators that are capable of detecting such excitations are labeled in legend.
}
\label{schematic_fig}
\end{figure}

In the N\'eel AFM phase, the system spontaneously breaks the continuous O(3) symmetry and selects a particular direction in spin space. 
As a consequence, the low-energy magnetic excitations consist of two linearly dispersive Goldstone modes corresponding to the restoration of the O(3) continuous symmetry. 
For example, if the spins are ordered along the $x$ direction
(denoted as ``xAF''), the two Goldstone modes correspond to generators $L^{12}$ and $L^{13}$ of the $\rm{SO(3)}$ group, see Figure \ref{schematic_fig}. 
The Goldstone modes are two-fold degenerate protected by the $\mathcal{PT}$ symmetry and cannot be directly read off from the spectra.
In addition, there is a gapped Higgs mode corresponding the amplitude fluctuations of the antiferromagnetic order parameter $\boldsymbol{N}$. 

The Goldstone modes arising from the transverse fluctuations of the N\'eel order parameter can be probed by the spin correlation functions $G_s$. 
As shown in Figure \ref{spin}(a) and (d), both the $J$-$Q_3$ and the $J_1$-$J_2$ model presents gapless excitations at the point $M=(\pi,\pi)$, with diverging spectral weight at low energies. 
Note that there also present additional gapless excitations at other momenta, but with vanishing low-energy spectral weight. These excitations come from the band folding of the Goldstone modes at the $M$ point and cannot be regarded as additional Goldstone modes~\footnote{For example, the gapless excitations at $\Gamma = (0,0)$ comes from the BZ folding due to the enlarged magnetic unit cell, and the gapless excitation at $X_2 = (0,\pi)$ that specifically appears in the $J_1$–$J_2$ model comes from the BZ folding due to the enlarged periodicity of the model.}. 

The Higgs mode as the amplitude fluctuations of the Néel order parameter is gapped within the  Néel AFM phase, and is in principle, also detectable at the $M$ point in spin channel $G_s$. However, the presence of Goldstone mode that appears at lower energies obscures its observation. 
Moreover, in two spatial dimension the Higgs excitation is strongly damped by the decay channel into two Goldstone modes, making it an ill-defined quasi-particle in the spin channel. 
Nevertheless, it was shown that the Higgs mode becomes well-defined quasi-particle in the bond correlation channel $G_b$, and could be captured in this channel without contamination from the low-lying Goldstone modes~\cite{sachdev1999universal,zwerger2004anomalous,dupuis2011infrared,podolsky2011visibility,podolsky2012spectral,gazit2013fate,gazit2013dynamics}. By matching the momentum conservation, such Higgs mode should appear at the $\Gamma$ point in the bond correlation channel $G_b$~\cite{podolsky2011visibility,qin2017amplitude,lohofer2015dynamical}. 
{In the bond correlation spectra (Fig. \ref{bond}), we clearly observe the Higgs excitation at $X_1$ for the $J_1$–$J_2$ model.} 
{However, for the $J$–$Q_3$ model we find vanishingly small low-energy spectral weight near the $\Gamma$ point. This makes it difficult to extract the gap of the longitudinal Higgs mode (see Supplementary Materials (SM)).
}

Additionally, in the bond correlation channel, we also identify a gapped excitation at $X_1=(\pi,0)$, see Figure \ref{bond}(a). 
Noticing that this mode only appears in the bond channel and is not visible in the spin channel, 
hence we assign this mode as the VBS fluctuation $V_x$ above the N\'eel phase. Mathematically, this mode is also the transverse fluctuation of the order parameter $\phi$ that corresponds to the SO(5) generator $L^{\alpha4}$ ($\alpha=1,2,3$) depending on the ordering direction of the AFM phase (see Table \ref{tab:example}). 
This excitation is gapped in the AFM phase, but should become gapless upon approaching the transition point $q_c$ once the emergent SO(5) symmetry is present. This aspect will be discussed in the following.

\textit{\color{blue}{Non-magnetic phases}.---}
Here we discuss the situations when the system lies deep within non-magnetic phases, namely the $\mathbb Z_4$ VBS phase of the $J$-$Q_3$ model and the trivial dimerized phase of the $J_1$-$J_2$ model. 
Since that the ground states break no continuous symmetries, all excitations are expected to be gapped. 
The fluctuations of the N\'eel order parameter comprise the three-fold triplet modes protected by the O(3) spin-rotational symmetry, as clearly seen in the spin excitation spectra Figure \ref{spin} (c) and (f). 
Such spin fluctuations above the VBS states are identified as the transverse fluctuations of the 5-component order parameter $\boldsymbol\phi$, and correspond to the SO(5) generators $L^{1\beta}$, $L^{2\beta}$ and $L^{3\beta}$ ($\beta=4,5$) depending on the ordering of the VBS state, see Table \ref{tab:example}.

In the bond correlation channel, we additionally observe low-lying spin-singlet excitations at momentum $X_1$ for the $J$-$Q_3$ model, as shown in Figure~\ref{bond}(c). These low-energy excitations can be traced to two different types of fluctuations. The first arises from the $V_x$ transverse VBS fluctuation above the yVBS state, which corresponds to the SO(5) generator $L^{45}$. Meanwhile, the second mode originates from the longitudinal fluctuation of the xVBS order parameter $V_x$, which also appears at the same momentum $X_1$. 
Naively, these two modes have distinct energies and should both be observable in our data. 
In practice, however, the longitudinal mode that locates at higher energies is also strongly damped by the decay process into two $L^{45}$ transverse modes (similar to the Higgs mode in the AFM phase), hence
requires much higher quality imaginary time correlation data to separately resolve the longitudinal $V_x$ excitation from the transverse $L^{45}$ mode. 
As a result, only the transverse mode that locates at lower energy could be clearly observable in our data, as shown in Figure~\ref{bond} (c). 
This result indicates that both the transverse and longitudinal VBS fluctuations are gapped in the VBS phase.


\textit{\color{blue}Quantum phase transition.---}
Here we discuss the evolution of excitation spectra across the QPT. 
For the $J_1$-$J_2$ model, this QPT is described by the O(3) Wilson-Fisher CFT. When approaching the QPT from the N\'eel phase, the two Goldstone modes remain gapless, while the Higgs mode becomes softened. Until one reaches the QPT point at $g_c$, the Higgs mode also becomes gapless, and it merges with the two Goldstone modes into three critical modes of the O(3) CFT. With further increase of $g$, the system transitions to the dimerized phase, and the excitations open a gap that corresponds to the three-fold triplet excitations. 

\begin{table}[thb]
\centering
\begin{tabular}{ccccc}
\hline
\hline
channel &  $\boldsymbol{k}$&  mode&gap \\
\hline
xAF $S^z$ &  $(\pi,\pi)$ &  $L^{13}$ & gapless \\
yAF $S^z$ &  $(\pi,\pi)$ &  $L^{23}$ & gapless \\
xAF $B_x$ &  $(\pi,0)$  & $L^{14}$ & gapless only at $q_c$\\
yAF $B_x$ &  $(\pi,0)$  & $L^{24}$ & gapless only at $q_c$\\
zAF $B_x$ &  $(\pi,0)$  & $L^{34}$ & gapless only at $q_c$\\
xVBS $S^z$ &  $(\pi,\pi)$  & $L^{34}$ & gapless only at $q_c$\\
yVBS $S^z$ &  $(\pi,\pi)$  & $L^{35}$ & gapless only at $q_c$\\
yVBS $B_x$ &  $(\pi,0)$  & $L^{45}$ & gapless only at $q_c$\cite{table1}\\

\hline
\hline
\end{tabular}
\caption{Identification between the operator channels and the low-energy transverse excitations in $J$-$Q_3$ model. The transverse excitations are labeled by the corresponding SO(5) generators.
}
\label{tab:example}
\end{table}


For the $J$–$Q_3$ model, as the system approaches the DQCP from the AFM phase (without losing generalities we assume that the system orders in the xAF state), the two Goldstone modes $L^{12}$ and $L^{13}$ that correspond to O(3) spin symmetry breaking remain gapless throughout. 
Meanwhile, the $V_x$ VBS fluctuation, $L^{14}$, that are gapped in the xAF phase become gapless at the DQCP, as shown at $X_1$ in Figure \ref{bond} (b).
By symmetry, we know that the $V_y$ VBS fluctuation, $L^{15}$, also becomes gapless at the DQCP. 
In conclusion, four gapless transverse modes appear at the AFM side of the DQCP.


The presence of four gapless transverse mode at DQCP also remains valid from the VBS side. 
It is well known that the VBS side host two emergent length scales near the DQCP: the smaller scale correspond to the rank-2 symmetric mass $(\phi_1^2+\phi_2^2+\phi_3^2-\phi_4^2-\phi_5^2)$ that controls the N\'eel-VBS transition, while the larger one corresponds to the quadrupled monopole term $(\phi_4+i\phi_5)^4+h.c.$ that breaks the U(1) topological symmetry down to $\mathbb Z_4$ \cite{tanaka_hu}. These two length scales correspond to two different gaps \cite{senthil2004deconfined}. The former corresponds to the triplet gap that locates at higher energy, while the latter corresponds to the pseudo-Goldstone mode $L^{45}$ that locates at lower energy.
As the system evolves from the yVBS phase to the DQCP, 
the triplet modes $L^{15}$, $L^{25}$, and $L^{35}$ exhibit a gap closing at the DQCP, as shown at the $M$ point in  the spin correlation [Figure~\ref{spin}(b) and (c)]. 
In addition, the transverse VBS fluctuation $L^{45}$, which is gapped inside the yVBS phase, quickly becomes softened approaching the DQCP and becomes gapless at the DQCP. 
The three triplet modes together with the VBS fluctuation constitutes the four Goldstone modes at the VBS side of DQCP. 

{The spectral gap of the longitudinal mode at the DQCP has important implications that determine the nature of this phase transition: 
if the longitudinal fluctuation becomes gapless, such ``DQCP'' would be a genuine QCP, where the longitudinal mode will merge with the four Goldstone modes to form the five critical modes of the SO(5) CFT; 
if the longitudinal fluctuation presents a finite gap, the DQCP would be a SO(5) symmetry-enhanced first-order transition, where its spontaneous breaking to O(4) would give rise to the four transverse Goldstone modes observed in our spectra. 
Previous studies have suggested that such ``DQCP'' to be weakly first-order, indicating that the longitudinal mode gap would be finite but extremely small. This put a crucial challenge to probe the gap of the longitudinal mode. 
Moreover, in our QMC simulations we find it difficult to probe the longitudinal mode from both AFM and VBS sides: 
The longitudinal fluctuation of the AFM phase should present at the $\Gamma$ point in the bond correlation. However, our measurement reveals vanishingly small spectral weight that makes it difficult to extract this gap. 
Meanwhile, the longitudinal mode of the VBS order is 
mixed with the transverse VBS fluctuation that appears at the same momenta, also making it difficult to separate from the low-lying transverse modes. 
As a result, we cannot draw definite conclusion on the precise nature of this phase transition from our spectral study alone. 
Nevertheless, considering that considerable numerical studies have reported that the phase transition of the $J$-$Q$ model does not align with a genuine QCP, we conclude that our spectral results are more consistent with the scenario of SO(5) symmetry-enhanced first-order transition that is spontaneous broken to O(4).
}

\textit{\color{blue} Conclusion.---}
We have investigated the spectral functions of both the $J$-$Q_3$ model and the $J_1$-$J_2$ model. In the $J_1$-$J_2$ model, two gapless magnon modes are observed in the N\'eel phase,  and the Higgs mode becomes gapless at the quantum critical point. The presence of three gapless critical modes at the QCP indicates that the system restores full $\rm{O(3)}$ symmetry. In the singlet phase, all low-energy modes are gapped.
In the $J$-$Q_3$ model, we similarly observe two gapless Goldstone modes in the N\'eel phase. 
At DQCP, in addition to the gapless transverse N\'eel fluctuation modes, the transverse fluctuations of the VBS order parameters also become gapless.
This indicates that the system exhibits an emergent $\mathrm{SO}(5)$ symmetry in the order parameter space, which unifies the five components of $\phi$.
The four Goldstone modes we found in the spectra are consistent with the recent result in which the scaling of entanglement entropy reveals the $SO(5)$ symmetry breaking.

\textit{\color{blue}Acknowledgment.---}
S.L., Y.L. and C.Z. contribute equally in this project.
This work is supported by the National Key Research and Development Program of China Grant No. 2022YFA1404204, and the National Natural Science Foundation of China Grant Nos. 12274086 and 12564021, and the Quantum Science and Technology-National Science and Technology Major Project (Grant No. 2024ZD0300104). Z.W. and Z.Y. are supported by the Scientific Research Project (No.WU2024B027) and the Start-up Funding of Westlake University. The authors thank the high-performance computing centers of Westlake University and the Beijing PARATERA Tech Co.,Ltd. for providing HPC resources.

\bibliography{JQ_SMA} 

@article{senthil2004deconfined,
  title     = {Deconfined quantum critical points},
  author    = {Senthil, Todadri and Vishwanath, Ashvin and Balents, Leon and Sachdev, Subir and Fisher, Matthew PA},
  journal   = {Science},
  volume    = {303},
  number    = {5663},
  pages     = {1490--1494},
  year      = {2004},
  publisher = {American Association for the Advancement of Science}
}

@article{senthil2004quantum,
  title     = {Quantum criticality beyond the Landau-Ginzburg-Wilson paradigm},
  author    = {Senthil, T and Balents, Leon and Sachdev, Subir and Vishwanath, Ashvin and Fisher, Matthew PA},
  journal   = {Physical Review B—Condensed Matter and Materials Physics},
  volume    = {70},
  number    = {14},
  pages     = {144407},
  year      = {2004},
  publisher = {APS}
}

@article{zhou2024so,
  title     = {SO (5) deconfined phase transition under the fuzzy-sphere microscope: Approximate conformal symmetry, pseudo-criticality, and operator spectrum},
  author    = {Zhou, Zheng and Hu, Liangdong and Zhu, W and He, Yin-Chen},
  journal   = {Physical Review X},
  volume    = {14},
  number    = {2},
  pages     = {021044},
  year      = {2024},
  publisher = {APS}
}

@article{shao2023progress,
  title     = {Progress on stochastic analytic continuation of quantum Monte Carlo data},
  author    = {Shao, Hui and Sandvik, Anders W},
  journal   = {Physics Reports},
  volume    = {1003},
  pages     = {1--88},
  year      = {2023},
  publisher = {Elsevier}
}

@article{wang2025extracting,
  title   = {Extracting the singularity of the logarithmic partition function},
  author  = {Wang, Zhe and Ding, Yi-Ming and Liu, Zenan and Yan, Zheng},
  journal = {arXiv preprint arXiv:2506.16111},
  year    = {2025}
}

@article{yan2019Dimer,
  title     = {Sweeping cluster algorithm for quantum spin systems with strong geometric restrictions},
  author    = {Yan, Zheng and Wu, Yongzheng and Liu, Chenrong and Sylju\aa{}sen, Olav F. and Lou, Jie and Chen, Yan},
  journal   = {Phys. Rev. B},
  volume    = {99},
  issue     = {16},
  pages     = {165135},
  numpages  = {6},
  year      = {2019},
  month     = {Apr},
  publisher = {American Physical Society},
  doi       = {10.1103/PhysRevB.99.165135},
  url       = {https://link.aps.org/doi/10.1103/PhysRevB.99.165135}
}

@article{Zhou2021amplitude,
  title     = {Amplitude Mode in Quantum Magnets via Dimensional Crossover},
  author    = {Zhou, Chengkang and Yan, Zheng and Wu, Han-Qing and Sun, Kai and Starykh, Oleg A. and Meng, Zi Yang},
  journal   = {Phys. Rev. Lett.},
  volume    = {126},
  issue     = {22},
  pages     = {227201},
  numpages  = {6},
  year      = {2021},
  month     = {Jun},
  publisher = {American Physical Society},
  doi       = {10.1103/PhysRevLett.126.227201},
  url       = {https://link.aps.org/doi/10.1103/PhysRevLett.126.227201}
}

@article{yan2021topological,
  title    = {Topological phase transition and single/multi anyon dynamics of ${Z}_2$ spin liquid},
  author   = {Yan, Zheng and Wang, Yan-Cheng and Ma, Nvsen and Qi, Yang and Meng, Zi Yang},
  journal  = {npj Quantum Mater.},
  year     = {2021},
  volume   = {6},
  numpages = {6},
  issue    = {6},
  pages    = {39},
  doi      = {10.1038/s41535-021-00338-1},
  url      = {https://doi.org/10.1038/s41535-021-00338-1}
}

@article{yan2021DimerImproved,
  title     = {Global scheme of sweeping cluster algorithm to sample among topological sectors},
  author    = {Yan, Zheng},
  journal   = {Phys. Rev. B},
  volume    = {105},
  issue     = {18},
  pages     = {184432},
  numpages  = {9},
  year      = {2022},
  month     = {May},
  publisher = {American Physical Society},
  doi       = {10.1103/PhysRevB.105.184432},
  url       = {https://link.aps.org/doi/10.1103/PhysRevB.105.184432}
}

@misc{wang2025probing,
  title         = {Probing phase transition and underlying symmetry breaking via entanglement entropy scanning},
  author        = {Zhe Wang and Zehui Deng and Zenan Liu and Zhiyan Wang and Yi-Ming Ding and Long Zhang and Wenan Guo and Zheng Yan},
  year          = {2025},
  eprint        = {2409.09942},
  archiveprefix = {arXiv},
  primaryclass  = {cond-mat.str-el},
  url           = {https://arxiv.org/abs/2409.09942}
}

@article{song2025evolution,
  title     = {Evolution of entanglement entropy at SU (N) deconfined quantum critical points},
  author    = {Song, Menghan and Zhao, Jiarui and Cheng, Meng and Xu, Cenke and Scherer, Michael and Janssen, Lukas and Meng, Zi Yang},
  journal   = {Science Advances},
  volume    = {11},
  number    = {6},
  pages     = {eadr0634},
  year      = {2025},
  publisher = {American Association for the Advancement of Science}
}

@article{deng2023improved,
  title     = {Improved scaling of the entanglement entropy of quantum antiferromagnetic Heisenberg systems},
  author    = {Deng, Zehui and Liu, Lu and Guo, Wenan and Lin, HQ},
  journal   = {Physical Review B},
  volume    = {108},
  number    = {12},
  pages     = {125144},
  year      = {2023},
  publisher = {APS}
}

@article{song2024extracting,
  title   = {Extracting subleading corrections in entanglement entropy at quantum phase transitions},
  author  = {Song, Menghan and Zhao, Jiarui and Meng, Zi Yang and Xu, Cenke and Cheng, Meng},
  journal = {SciPost Physics},
  volume  = {17},
  number  = {1},
  pages   = {010},
  year    = {2024}
}

@article{sandvik2007evidence,
  title     = {Evidence for Deconfined Quantum Criticality in a Two-Dimensional Heisenberg Model with Four-Spin Interactions},
  author    = {Sandvik, Anders W},
  journal   = {Physical review letters},
  volume    = {98},
  number    = {22},
  pages     = {227202},
  year      = {2007},
  publisher = {APS}
}

@article{nahum2015deconfined,
  title     = {Deconfined quantum criticality, scaling violations, and classical loop models},
  author    = {Nahum, Adam and Chalker, JT and Serna, P and Ortu{\~n}o, M and Somoza, AM},
  journal   = {Physical Review X},
  volume    = {5},
  number    = {4},
  pages     = {041048},
  year      = {2015},
  publisher = {APS}
}

@article{qin2017duality,
  title     = {Duality between the deconfined quantum-critical point and the bosonic topological transition},
  author    = {Qin, Yan Qi and He, Yuan-Yao and You, Yi-Zhuang and Lu, Zhong-Yi and Sen, Arnab and Sandvik, Anders W and Xu, Cenke and Meng, Zi Yang},
  journal   = {Physical Review X},
  volume    = {7},
  number    = {3},
  pages     = {031052},
  year      = {2017},
  publisher = {APS}
}

@article{wang2017deconfined,
  title     = {Deconfined quantum critical points: symmetries and dualities},
  author    = {Wang, Chong and Nahum, Adam and Metlitski, Max A and Xu, Cenke and Senthil, T},
  journal   = {Physical Review X},
  volume    = {7},
  number    = {3},
  pages     = {031051},
  year      = {2017},
  publisher = {APS}
}

@article{sandvik2010continuous,
  title     = {Continuous Quantum Phase Transition between an Antiferromagnet and a Valence-Bond Solid in Two Dimensions: Evidence for Logarithmic Corrections to Scaling},
  author    = {Sandvik, Anders W},
  journal   = {Physical review letters},
  volume    = {104},
  number    = {17},
  pages     = {177201},
  year      = {2010},
  publisher = {APS}
}

@article{nahum2015emergent,
  title     = {Emergent SO (5) symmetry at the N{\'e}el to valence-bond-solid transition},
  author    = {Nahum, Adam and Serna, P and Chalker, JT and Ortu{\~n}o, M and Somoza, AM},
  journal   = {Physical review letters},
  volume    = {115},
  number    = {26},
  pages     = {267203},
  year      = {2015},
  publisher = {APS}
}

@article{sreejith2019emergent,
  title     = {Emergent SO (5) symmetry at the columnar ordering transition in the classical cubic dimer model},
  author    = {Sreejith, GJ and Powell, Stephen and Nahum, Adam},
  journal   = {Physical review letters},
  volume    = {122},
  number    = {8},
  pages     = {080601},
  year      = {2019},
  publisher = {APS}
}

@article{deng2024diagnosing,
  title     = {Diagnosing quantum phase transition order and deconfined criticality via entanglement entropy},
  author    = {Deng, Zehui and Liu, Lu and Guo, Wenan and Lin, Hai-Qing},
  journal   = {Physical Review Letters},
  volume    = {133},
  number    = {10},
  pages     = {100402},
  year      = {2024},
  publisher = {APS}
}

@article{kuklov2008deconfined,
  title     = {Deconfined criticality: Generic first-order transition in the SU (2) symmetry case},
  author    = {Kuklov, AB and Matsumoto, M and Prokof’Ev, NV and Svistunov, BV and Troyer, M},
  journal   = {Physical review letters},
  volume    = {101},
  number    = {5},
  pages     = {050405},
  year      = {2008},
  publisher = {APS}
}

@article{chen2013deconfined,
  title     = {Deconfined criticality flow in the Heisenberg model with ring-exchange interactions},
  author    = {Chen, Kun and Huang, Yuan and Deng, Youjin and Kuklov, AB and Prokof’ev, NV and Svistunov, BV},
  journal   = {Physical review letters},
  volume    = {110},
  number    = {18},
  pages     = {185701},
  year      = {2013},
  publisher = {APS}
}

@article{nakayama2016necessary,
  title     = {Necessary condition for emergent symmetry from the conformal bootstrap},
  author    = {Nakayama, Yu and Ohtsuki, Tomoki},
  journal   = {Physical Review Letters},
  volume    = {117},
  number    = {13},
  pages     = {131601},
  year      = {2016},
  publisher = {APS}
}

@article{li2022bootstrapping,
  title     = {Bootstrapping conformal QED3 and deconfined quantum critical point},
  author    = {Li, Zhijin},
  journal   = {Journal of High Energy Physics},
  volume    = {2022},
  number    = {11},
  pages     = {1--20},
  year      = {2022},
  publisher = {Springer}
}

@article{poland2019conformal,
  title     = {The conformal bootstrap: Theory, numerical techniques, and applications},
  author    = {Poland, David and Rychkov, Slava and Vichi, Alessandro},
  journal   = {Reviews of Modern Physics},
  volume    = {91},
  number    = {1},
  pages     = {015002},
  year      = {2019},
  publisher = {APS}
}

@article{nahum2020note,
  title     = {Note on Wess-Zumino-Witten models and quasiuniversality in 2+ 1 dimensions},
  author    = {Nahum, Adam},
  journal   = {Physical Review B},
  volume    = {102},
  number    = {20},
  pages     = {201116},
  year      = {2020},
  publisher = {APS}
}

@article{ma2020theory,
  title     = {Theory of deconfined pseudocriticality},
  author    = {Ma, Ruochen and Wang, Chong},
  journal   = {Physical Review B},
  volume    = {102},
  number    = {2},
  pages     = {020407},
  year      = {2020},
  publisher = {APS}
}

@article{zhao2020multicritical,
  title     = {Multicritical deconfined quantum criticality and Lifshitz point of a helical valence-bond phase},
  author    = {Zhao, Bowen and Takahashi, Jun and Sandvik, Anders W},
  journal   = {Physical Review Letters},
  volume    = {125},
  number    = {25},
  pages     = {257204},
  year      = {2020},
  publisher = {APS}
}

@article{chen2024phases,
  title     = {Phases of (2+ 1) d so (5) nonlinear sigma model with a topological term on a sphere: multicritical point and disorder phase},
  author    = {Chen, Bin-Bin and Zhang, Xu and Wang, Yuxuan and Sun, Kai and Meng, Zi Yang},
  journal   = {Physical Review Letters},
  volume    = {132},
  number    = {24},
  pages     = {246503},
  year      = {2024},
  publisher = {APS}
}

@article{lu2021self,
  title     = {Self-duality protected multicriticality in deconfined quantum phase transitions},
  author    = {Lu, Da-Chuan and Xu, Cenke and You, Yi-Zhuang},
  journal   = {Physical Review B},
  volume    = {104},
  number    = {20},
  pages     = {205142},
  year      = {2021},
  publisher = {APS}
}

@article{chester2024bootstrapping,
  title     = {Bootstrapping deconfined quantum tricriticality},
  author    = {Chester, Shai M and Su, Ning},
  journal   = {Physical Review Letters},
  volume    = {132},
  number    = {11},
  pages     = {111601},
  year      = {2024},
  publisher = {APS}
}

@article{wang2022scaling,
  title   = {Scaling of the disorder operator at deconfined quantum criticality},
  author  = {Wang, Yan-Cheng and Ma, Nvsen and Cheng, Meng and Meng, Zi Yang},
  journal = {SciPost Physics},
  volume  = {13},
  number  = {6},
  pages   = {123},
  year    = {2022}
}

@article{lou2009antiferromagnetic,
  title     = {Antiferromagnetic to valence-bond-solid transitions in two-dimensional SU (N) Heisenberg models with multispin interactions},
  author    = {Lou, Jie and Sandvik, Anders W and Kawashima, Naoki},
  journal   = {Physical Review B—Condensed Matter and Materials Physics},
  volume    = {80},
  number    = {18},
  pages     = {180414},
  year      = {2009},
  publisher = {APS}
}

@article{melko2008scaling,
  title     = {Scaling in the fan of an unconventional quantum critical point},
  author    = {Melko, Roger G and Kaul, Ribhu K},
  journal   = {Physical review letters},
  volume    = {100},
  number    = {1},
  pages     = {017203},
  year      = {2008},
  publisher = {APS}
}

@article{jiang2008antiferromagnet,
  title     = {From an antiferromagnet to a valence bond solid: evidence for a first-order phase transition},
  author    = {Jiang, Fu-Jiun and Nyfeler, Matthias and Chandrasekharan, S and Wiese, Uwe-Jens},
  journal   = {Journal of Statistical Mechanics: Theory and Experiment},
  volume    = {2008},
  number    = {02},
  pages     = {P02009},
  year      = {2008},
  publisher = {IOP Publishing}
}

@article{dalla2015fractional,
  title     = {Fractional excitations in the square-lattice quantum antiferromagnet},
  author    = {Dalla Piazza, Bastien and Mourigal, M and Christensen, Niels Bech and Nilsen, GJ and Tregenna-Piggott, P and Perring, TG and Enderle, Mechtild and McMorrow, Desmond Francis and Ivanov, DA and R{\o}nnow, Henrik Moodysson},
  journal   = {Nature physics},
  volume    = {11},
  number    = {1},
  pages     = {62--68},
  year      = {2015},
  publisher = {Nature Publishing Group UK London}
}

@article{song2023dynamical,
  title     = {Dynamical properties of quantum many-body systems with long-range interactions},
  author    = {Song, Menghan and Zhao, Jiarui and Zhou, Chengkang and Meng, Zi Yang},
  journal   = {Physical Review Research},
  volume    = {5},
  number    = {3},
  pages     = {033046},
  year      = {2023},
  publisher = {APS}
}

@article{sandvik1991quantum,
  title     = {Quantum Monte Carlo simulation method for spin systems},
  author    = {Sandvik, Anders W and Kurkij{\"a}rvi, Juhani},
  journal   = {Physical Review B},
  volume    = {43},
  number    = {7},
  pages     = {5950},
  year      = {1991},
  publisher = {APS}
}

@article{Sandvik1999SSE,
  title     = {Stochastic series expansion method with operator-loop update},
  author    = {Sandvik, Anders W.},
  journal   = {Phys. Rev. B},
  volume    = {59},
  issue     = {22},
  pages     = {R14157--R14160},
  numpages  = {0},
  year      = {1999},
  month     = {Jun},
  publisher = {American Physical Society},
  doi       = {10.1103/PhysRevB.59.R14157},
  url       = {https://link.aps.org/doi/10.1103/PhysRevB.59.R14157}
}

@article{syljuaasen2002quantum,
  title     = {Quantum Monte Carlo with directed loops},
  author    = {Sylju{\aa}sen, Olav F and Sandvik, Anders W},
  journal   = {Physical Review E},
  volume    = {66},
  number    = {4},
  pages     = {046701},
  year      = {2002},
  publisher = {APS}
}

@article{sandvik2019stochastic,
  title   = {Stochastic series expansion methods},
  author  = {Sandvik, Anders W},
  journal = {arXiv preprint arXiv:1909.10591},
  year    = {2019}
}

@article{desai2021resummation,
  title     = {Resummation-based quantum monte carlo for quantum paramagnetic phases},
  author    = {Desai, Nisheeta and Pujari, Sumiran},
  journal   = {Physical Review B},
  volume    = {104},
  number    = {6},
  pages     = {L060406},
  year      = {2021},
  publisher = {APS}
}

@article{zhao2022scaling,
  title     = {Scaling of entanglement entropy at deconfined quantum criticality},
  author    = {Zhao, Jiarui and Wang, Yan-Cheng and Yan, Zheng and Cheng, Meng and Meng, Zi Yang},
  journal   = {Physical Review Letters},
  volume    = {128},
  number    = {1},
  pages     = {010601},
  year      = {2022},
  publisher = {APS}
}

@article{ma2018anomalous,
  title     = {Anomalous quantum-critical scaling corrections in two-dimensional antiferromagnets},
  author    = {Ma, Nvsen and Weinberg, Phillip and Shao, Hui and Guo, Wenan and Yao, Dao-Xin and Sandvik, Anders W},
  journal   = {Physical Review Letters},
  volume    = {121},
  number    = {11},
  pages     = {117202},
  year      = {2018},
  publisher = {APS}
}

@article{lohofer2015dynamical,
  title     = {Dynamical structure factors and excitation modes of the bilayer Heisenberg model},
  author    = {Loh{\"o}fer, M and Coletta, T and Joshi, DG and Assaad, FF and Vojta, M and Wessel, S and Mila, F},
  journal   = {Physical Review B},
  volume    = {92},
  number    = {24},
  pages     = {245137},
  year      = {2015},
  publisher = {APS}
}

@article{qin2017amplitude,
  title     = {Amplitude mode in three-dimensional dimerized antiferromagnets},
  author    = {Qin, Yan Qi and Normand, Bruce and Sandvik, Anders W and Meng, Zi Yang},
  journal   = {Physical review letters},
  volume    = {118},
  number    = {14},
  pages     = {147207},
  year      = {2017},
  publisher = {APS}
}

@article{sandvik1998stochastic,
  title     = {Stochastic method for analytic continuation of quantum Monte Carlo data},
  author    = {Sandvik, Anders W},
  journal   = {Physical Review B},
  volume    = {57},
  number    = {17},
  pages     = {10287},
  year      = {1998},
  publisher = {APS}
}

@article{sandvik2016constrained,
  title     = {Constrained sampling method for analytic continuation},
  author    = {Sandvik, Anders W},
  journal   = {Physical Review E},
  volume    = {94},
  number    = {6},
  pages     = {063308},
  year      = {2016},
  publisher = {APS}
}

@article{beach2004identifying,
  title   = {Identifying the maximum entropy method as a special limit of stochastic analytic continuation},
  author  = {Beach, KSD},
  journal = {arXiv preprint cond-mat/0403055},
  year    = {2004}
}

@article{podolsky2012spectral,
  title     = {Spectral functions of the Higgs mode near two-dimensional quantum critical points},
  author    = {Podolsky, Daniel and Sachdev, Subir},
  journal   = {Physical Review B—Condensed Matter and Materials Physics},
  volume    = {86},
  number    = {5},
  pages     = {054508},
  year      = {2012},
  publisher = {APS}
}

@article{gazit2013fate,
  title     = {Fate of the Higgs mode near quantum criticality},
  author    = {Gazit, Snir and Podolsky, Daniel and Auerbach, Assa},
  journal   = {Physical Review Letters},
  volume    = {110},
  number    = {14},
  pages     = {140401},
  year      = {2013},
  publisher = {APS}
}

@article{gazit2013dynamics,
  title     = {Dynamics and conductivity near quantum criticality},
  author    = {Gazit, Snir and Podolsky, Daniel and Auerbach, Assa and Arovas, Daniel P},
  journal   = {Physical Review B—Condensed Matter and Materials Physics},
  volume    = {88},
  number    = {23},
  pages     = {235108},
  year      = {2013},
  publisher = {APS}
}

@article{zhao2019symmetry,
  title     = {Symmetry-enhanced discontinuous phase transition in a two-dimensional quantum magnet},
  author    = {Zhao, Bowen and Weinberg, Phillip and Sandvik, Anders W},
  journal   = {Nature Physics},
  volume    = {15},
  number    = {7},
  pages     = {678--682},
  year      = {2019},
  publisher = {Nature Publishing Group UK London}
}

@article{takahashi2020valence,
  title     = {Valence-bond solids, vestigial order, and emergent SO (5) symmetry in a two-dimensional quantum magnet},
  author    = {Takahashi, Jun and Sandvik, Anders W},
  journal   = {Physical Review Research},
  volume    = {2},
  number    = {3},
  pages     = {033459},
  year      = {2020},
  publisher = {APS}
}

@article{ma2018dynamical,
  title     = {Dynamical signature of fractionalization at a deconfined quantum critical point},
  author    = {Ma, Nvsen and Sun, Guang-Yu and You, Yi-Zhuang and Xu, Cenke and Vishwanath, Ashvin and Sandvik, Anders W and Meng, Zi Yang},
  journal   = {Physical Review B},
  volume    = {98},
  number    = {17},
  pages     = {174421},
  year      = {2018},
  publisher = {APS}
}

@article{ma2019role,
  title     = {Role of Noether’s theorem at the deconfined quantum critical point},
  author    = {Ma, Nvsen and You, Yi-Zhuang and Meng, Zi Yang},
  journal   = {Physical Review Letters},
  volume    = {122},
  number    = {17},
  pages     = {175701},
  year      = {2019},
  publisher = {APS}
}

@article{podolsky2011visibility,
  title     = {Visibility of the amplitude (Higgs) mode in condensed matter},
  author    = {Podolsky, Daniel and Auerbach, Assa and Arovas, Daniel P},
  journal   = {Physical Review B—Condensed Matter and Materials Physics},
  volume    = {84},
  number    = {17},
  pages     = {174522},
  year      = {2011},
  publisher = {APS}
}

@article{sachdev1999universal,
  title     = {Universal relaxational dynamics near two-dimensional quantum critical points},
  author    = {Sachdev, Subir},
  journal   = {Physical Review B},
  volume    = {59},
  number    = {21},
  pages     = {14054},
  year      = {1999},
  publisher = {APS}
}

@article{zwerger2004anomalous,
  title     = {Anomalous fluctuations in phases with a broken continuous symmetry},
  author    = {Zwerger, W},
  journal   = {Physical review letters},
  volume    = {92},
  number    = {2},
  pages     = {027203},
  year      = {2004},
  publisher = {APS}
}

@article{dupuis2011infrared,
  title     = {Infrared behavior in systems with a broken continuous symmetry: Classical O (N) model versus interacting bosons},
  author    = {Dupuis, Nicolas},
  journal   = {Physical Review E—Statistical, Nonlinear, and Soft Matter Physics},
  volume    = {83},
  number    = {3},
  pages     = {031120},
  year      = {2011},
  publisher = {APS}
}

@article{dorneich2001accessing,
  title     = {Accessing the dynamics of large many-particle systems using the stochastic series expansion},
  author    = {Dorneich, Ansgar and Troyer, Matthias},
  journal   = {Physical Review E},
  volume    = {64},
  number    = {6},
  pages     = {066701},
  year      = {2001},
  publisher = {APS}
}

@article{takahashi2024so,
  title   = {SO (5) multicriticality in two-dimensional quantum magnets},
  author  = {Takahashi, Jun and Shao, Hui and Zhao, Bowen and Guo, Wenan and Sandvik, Anders W},
  journal = {arXiv preprint arXiv:2405.06607},
  year    = {2024}
}

@article{matsumoto2001ground,
  title     = {Ground-state phase diagram of quantum Heisenberg antiferromagnets on the anisotropic dimerized square lattice},
  author    = {Matsumoto, Munehisa and Yasuda, Chitoshi and Todo, Synge and Takayama, Hajime},
  journal   = {Physical Review B},
  volume    = {65},
  number    = {1},
  pages     = {014407},
  year      = {2001},
  publisher = {APS}
}

@article{wenzel2009comprehensive,
  title     = {Comprehensive quantum Monte Carlo study of the quantum critical points in planar dimerized/quadrumerized Heisenberg models},
  author    = {Wenzel, Sandro and Janke, Wolfhard},
  journal   = {Physical Review B—Condensed Matter and Materials Physics},
  volume    = {79},
  number    = {1},
  pages     = {014410},
  year      = {2009},
  publisher = {APS}
}

@article{chakravarty1988low,
  title     = {Low-temperature behavior of two-dimensional quantum antiferromagnets},
  author    = {Chakravarty, Sudip and Halperin, Bertrand I and Nelson, David R},
  journal   = {Physical review letters},
  volume    = {60},
  number    = {11},
  pages     = {1057},
  year      = {1988},
  publisher = {APS}
}

@article{haldane19883,
  title     = {O (3) nonlinear $\sigma$ model and the topological distinction between integer-and half-integer-spin antiferromagnets in two dimensions},
  author    = {Haldane, F Duncan M},
  journal   = {Physical review letters},
  volume    = {61},
  number    = {8},
  pages     = {1029},
  year      = {1988},
  publisher = {APS}
}

@article{chubukov1994theory,
  title     = {Theory of two-dimensional quantum Heisenberg antiferromagnets with a nearly critical ground state},
  author    = {Chubukov, Andrey V and Sachdev, Subir and Ye, Jinwu},
  journal   = {Physical Review B},
  volume    = {49},
  number    = {17},
  pages     = {11919},
  year      = {1994},
  publisher = {APS}
}

@article{wenzel2008evidence,
  title     = {Evidence for an Unconventional Universality Class from a Two-Dimensional<? format?> Dimerized Quantum Heisenberg Model},
  author    = {Wenzel, Sandro and Bogacz, Leszek and Janke, Wolfhard},
  journal   = {Physical review letters},
  volume    = {101},
  number    = {12},
  pages     = {127202},
  year      = {2008},
  publisher = {APS}
}

@booklet{commutator,
  title = {The vanishing commutator $[ \sum_i S^z_i , \boldsymbol{S}_k \cdot \boldsymbol{S}_l ] = 0$ implies that the correlation function $G_s(\Gamma,\tau)$ remains identically zero for all imaginary times $\tau$. Consequently, the spin spectral function vanishes at the $\Gamma$ point.}
}

@booklet{bond_commutator,
  title = {The correlation function ${G_b(\Gamma,\infty)}$ remains nonzero for both models due to the antiferromagnetic interactions. {I}n the ${J_1}$–${J_2}$ model, the spatial translational symmetry is broken along the $x$ direction, leading to ${G_b(X_1,\infty)\neq 0}$. {I}n the VBS phase, the system spontaneously breaks the ${{Z}_4}$ symmetry, and likewise ${G_b(X_1,\infty)\neq 0}$. {T}he persistence of a nonzero correlation function at infinite imaginary time prevents us from obtaining a high-quality spectral function; {T}herefore, we do not present it in the figure.}
}

@misc{table1,
  key  = {This pseudo-Goldstone mode corresponds to the larger length scale away from DQCP,},
  note = {hence in our simulation this gap becomes visible only when the system lies sufficiently away
          from ${q_c}$.}
}

@article{chen2024spin,
  title   = {Spin excitations of the Shastry-Sutherland model--altermagnetism and proximate deconfined quantum criticality},
  author  = {Chen, Hongyu and Duan, Guijing and Liu, Changle and Cui, Yi and Yu, Weiqiang and Xie, ZY and Yu, Rong},
  journal = {arXiv preprint arXiv:2411.00301},
  year    = {2024}
}

@article{tanaka_hu,
  title     = {Many-Body Spin Berry Phases Emerging from the $\ensuremath{\pi}$-Flux State: Competition between Antiferromagnetism and the Valence-Bond-Solid State},
  author    = {Tanaka, Akihiro and Hu, Xiao},
  journal   = {Phys. Rev. Lett.},
  volume    = {95},
  issue     = {3},
  pages     = {036402},
  numpages  = {4},
  year      = {2005},
  month     = {Jul},
  publisher = {American Physical Society},
  doi       = {10.1103/PhysRevLett.95.036402},
  url       = {https://link.aps.org/doi/10.1103/PhysRevLett.95.036402}
}
\end{document}